\documentclass[aps,amsmath,amssymb,reprint,superscriptaddress]{revtex4-1}
\usepackage{graphicx}
\usepackage{dcolumn}
\usepackage{bm}
\usepackage{CJKutf8}
\usepackage{natbib}

\begin{document}
\title{Raman lasing in a hollow, bottle-like microresonator}
\author{Yuta Ooka}
\affiliation{Light-Matter Interactions Unit, Okinawa Institute of Science and Technology Graduate University, Onna, Okinawa 904-0495, Japan}
\affiliation{Dept. of Electronics and Electrical Engineering, Faculty of Science and Technology, Keio University, 3-14-1, Hiyoshi, Kohoku-ku, Yokohama 223-8522, Japan}
\author{Yong Yang}
\author{Jonathan Ward}
\author{S\'ile Nic Chormaic}
\affiliation{Light-Matter Interactions Unit, Okinawa Institute of Science and Technology Graduate University, Onna, Okinawa 904-0495, Japan}

\begin{abstract}
We report on the fabrication of an ultrahigh quality factor, bottle-like microresonator from a microcapillary, and the realization of Raman lasing therein at pump wavelengths of $1.55~\mathrm{\mu m}$ and $780~\mathrm{nm}$. The dependence of the Raman laser threshold on mode volume is investigated. The mode volume of the fundamental bottle mode is calculated and compared with that of a microsphere. Third-order cascaded Raman lasing was observed when pumped at $780~\mathrm{nm}$. In principle, Raman lasing in a hollow bottle-like microresonator can be used in sensing applications. As an example, we briefly discuss the possibility of a high dynamic range, high resolution aerostatic pressure sensor.
\end{abstract}
\maketitle

Whispering gallery mode (WGM) microresonators (WGRs) have high quality (Q) factors and relatively small mode volumes.  Based on these properties, they can be used for the demonstration and implementation of many nonlinear effects, such as stimulated Brillouin scattering\cite{Bahl2011}, four-wave mixing (FWM)\cite{Agha2009}, soliton generation\cite{Herr2014}, and Kerr optical switching\cite{Pollinger2010,Yoshiki2014}.  The observation of Raman lasing in WGRs is based on the resonator material's phonon band.  Thus, in comparison with Brillouin scattering or FWM, phase matching is automatically satisfied; this makes Raman lasing in WGRs relatively easy to implement. The ultrahigh optical Q in WGRs guarantees a very low Raman lasing threshold and even cascaded processes can be achieved, as has been demonstrated in silica microspheres\cite{Spillane2002,Min2003}, microtoroids\cite{Kippenberg2004, Lu2011}, chalcogenide microspheres\cite{Vanier2013,Vanier2014}, and PDMS (polymer) WGRs\cite{Li2013}. Raman lasing in  WGRs can be very useful for sensing applications since (i) it does not require a dopant in the resonator's material and (ii) it decreases the effective linewidth of the WGM  through Raman gain. Single nanoparticle detection using Raman lasing has been achieved\cite{Li2014, Ozdemir2014, Peng2014}, thereby illustrating the high detection resolution attainable.

Conventional WGRs, such as microspheres, microdisks, and microtoroids, confine the optical mode to narrow rings along the equator. Even though the mode volume is small, the tuning range is limited by the design. Bottle-like microresonators (BLMRs) were developed\cite{Pollinger2009} since they have good strain and stress tunability. The good tunability makes them attractive devices for sensing and a number of applications have recently been developed\cite{Sumetsky2010, Murugan2011}.
Instead of heating a tapered optical fiber to achieve a solid bottle shape\cite{Pollinger2009, Ward2006}, a silica (or other material) microcapillary is used to create a hollow bottle. This creates a highly prolate shape with an empty channel inside. The bottle mode is sensitive to the material contained within the microcapillary, so that this device is a promising candidate for optofluidic sensing; to date, refractive index sensing has been demonstrated\cite{Murugan2011, Zhang2014}.

As with their solid counterparts, hollow BLMRs can also be used for studying nonlinear behavior. The nonlinear response of a material is related to $Q^2/V$, where $V$ is the mode volume.  Ultra-high Q-factors of up to $10^8$ have been achieved in hollow BLMRs with wall thicknesses greater than 12 $\mu$m \cite{Bahl2013,Han2014} and relatively small mode volumes. Brillouin scattering was studied in these thick-walled BLMRs and the authors demonstrated that the devices could function as a viscosity sensors for the contained fluid \cite{Bahl2013}. Using stimulated Brillouin scattering, aerostatic pressure sensing has also been realized. In this case, a change in the internal pressure shifts the BLMRs mechanical mode, which is measured by homodyne detection\cite{Han2014}.

Here, we report on another nonlinear effect in hollow BLMRs, namely Raman lasing. We demonstrate low-threshold Raman lasing in a BLMR with a Q-factor as high as $10^8$. The threshold for the Raman lasing is higher than  that for a microsphere of a similar diameter due to the larger mode volume of the BLMR. As with other WGRs, we observe a cascaded Raman process  when the pump power is high enough.
\begin{figure}
	\begin{center}
     \includegraphics[width=65mm]{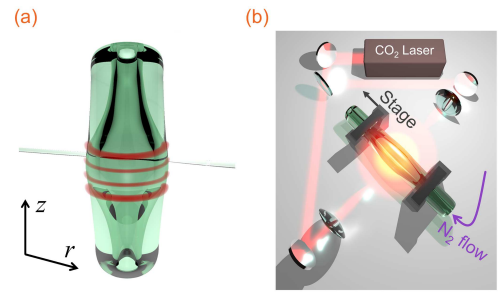}
    \end{center}
\caption{(a) Bottle whispering gallery mode resonator. The red lines represent the mode light traveling in the resonator. (b) Schematic of the CO$_2$ laser bottle-like resonator fabrication setup.}
\label{f0}
\end{figure}

\begin{figure}
	\begin{center}
	\includegraphics[width = 60mm]{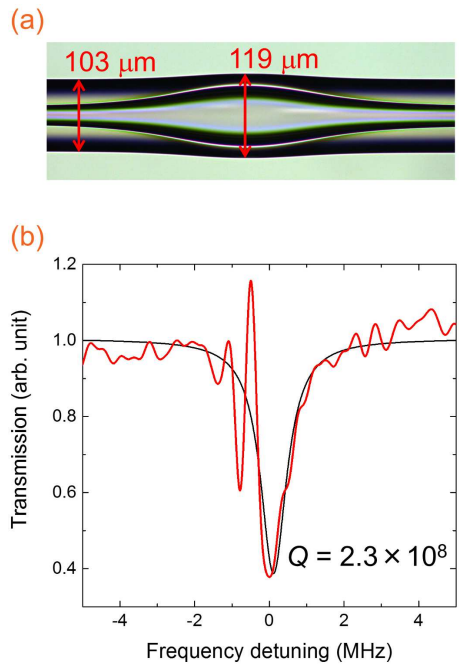}
	\end{center}
	\caption{(a) Microscope image of the hollow resonator. (b) Transmittance spectrum of a mode (red line) that was used for generating stimulated Raman scattering. The center wavelength of this mode is about $1545.09~\mathrm{nm}$. The transmission is normalized at the level of the laser background. $Q$ is $2.3\times10^8$ by Lorentzian fitting (black line).}
	\label{f1}
\end{figure}

A hollow BLMR (Fig.~\ref{f0}(a)) was fabricated using the setup illustrated in Fig.~\ref{f0}(b). A silica capillary, with an outer diameter of $350~\mathrm{\mu m}$ and an inner diameter of $100~\mathrm{\mu m}$, was clamped onto two translation stages. Counter-propagating CO$_2$ beams were focused on each side of the capillary and a CCD camera was used to monitor the fabrication process from the top. A weak CO$_2$ power was initially used in order to remove the polymer coating from the capillary. The power was then gently increased to soften the  capillary while the two stages pulled on both ends. Hence, a tapered capillary was fabricated with a decreasing diameter along its length.  Next, the tapered capillary was connected to a nitrogen gas source and the inner pressure of the capillary was kept at around 3~bar. Finally, the $\mathrm{CO_2}$ laser beams were focused back on the tapered zone. The molten wall of the tapered capillary  swelled gradually until the desired outer diameter was achieved. After fabrication, the capillary with the hollow BLMR was removed from the stages and glued to a glass holder. From the microscope image of the device in Fig.~\ref{f1}(a), it can be seen that the BLMR appears as a small bump along the microcapillary. The outer diameter of the bottle was $119~\mathrm{\mu m}$ and the diameter of the tapered capillary was $103~\mathrm{\mu m}$. The thickness of the BLMR wall was estimated to be $25~\mathrm{\mu m}$.

The BLMR was placed on a 3D nanopositioner. A tapered fiber was used to evanescently couple light from a tunable diode laser (Newfocus Velocity 6728) into the resonator. In the experiment, the coupling gap was adjusted to minimize the Raman lasing threshold. To maintain the coupling condition, the whole system was isolated from the environment using an enclosure. First, the pump laser was scanned over a range spanning 17.5 GHz, with a speed of $29~\mathrm{nm/s}$.  The pump laser was tuned from the blue to the red side of the optical mode and the transmission spectrum was recorded. To excite different WGMs, the BLMR was moved using the nanopositioner so that the tapered fiber traversed the equator from one side to the other. The position was changed until a ringing effect\cite{Trebaol2011} was observed. The ringing is an indication of an ultrahigh Q mode and it occurs as the light stored in the microresonator beats with the light from the scanned laser in the taper before the cavity mode decays by intrinsic loss. Figure~\ref{f1}(b) shows the mode used for observing Raman lasing. The measured loaded Q of this mode is $2.3\times10^8$.

\begin{figure}
	\begin{center}
	\includegraphics[width = 60mm]{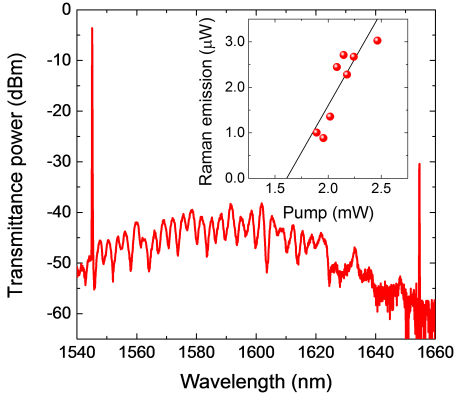}
	\end{center}
	\caption{Raman spectrum when the pump power is above threshold. The wavelength of the pump laser and Raman scattering are $1545.1~\mathrm{nm}$ and $1654.6~\mathrm{nm}$, respectively. Inset: relationship between the power of the pump and Raman scattering. The black line is a linear fit to the data.}
	\label{f2}
\end{figure}

Light from the taper output was split, with one portion sent to a photodiode and the other sent to an optical spectrum analyzer (Anritsu MS9740A). The resolution of the spectrum analyzer was set to 0.07~nm with a wavelength range from 1540~nm to 1660~nm. The laser frequency was tuned into resonance with the cavity mode. Due to the thermal and Kerr effects, when the laser is tuned from the blue side of the WGM, the WGM shifts in the same direction as the laser tuning so that the pump and the WGM become thermally locked. The input power was increased while the system was in this self-stabilizing state and stimulated Raman scattering (SRS), i.e., Raman lasing, was observed on the spectrum analyzer. Figure~\ref{f2} shows  an example of this effect. The strong line to the left is the pump laser with a wavelength of 1545.1~nm. The peak at 1654.5~nm is the Raman lasing, which appears when the pump laser power is high enough; this is the SRS of the BLMR. The interval between the pump and Raman peaks is about 110~nm, which corresponds to the 13.5~THz peak on the Raman gain spectrum of silica. By varying the input pump power, the Raman threshold was measured and is shown in the inset of Fig.~\ref{f2}. Note that the total transmission efficiency of the taper used for coupling was 87\% and the final measured threshold was about 1.6~mW.

The power threshold of SRS is given by\cite{Spillane2002} $P_{th}\propto {V}/{Q^2}$, where $Q$ represents the optical Q-factor of the pump mode and $V$ is the mode volume defined as:
\begin{equation}
V=\iiint \epsilon\frac{|E(r,\varphi,z)|^2}{|E_{max}(r,\varphi,z)|^2}rdrd\varphi dz.
\label{eq4}
\end{equation}
Here, $\epsilon$ is the permittivity of the medium, $E$ is the electromagnetic field amplitude of the BLMR mode and $E_{max}$ is the maximum of $E$.

The WGMs in a BLMR are different from the WGMs in a microsphere since the BLMR modes propagation constant has a component along the \textit{z}-axis (Fig.\ref{f0}(a))\cite{Sumetsky2004, Louyer2005}. Let us assume that the profile of the BLMR is parabolic with $R(z)=R_0(1-1/2(\zeta z)^2)$, where $R_0$ is the maximum radius and $\zeta^2$ is the curvature of the profile. The first order radial mode can be explicitly solved in cylindrical coordinates\cite{Sumetsky2004, Louyer2005, Zhang2014}, yielding the wave number of the mode as:
\begin{equation}
k_{lp}=\sqrt{\frac{l^2}{R_0^2}+(p+\frac{1}{2})\Gamma_l}.
\label{eq1}
\end{equation}
The mode is numbered using two indices, $l$ for the azimuthal and $p$ for the axial mode numbers. $p$ represents the number of maxima along the \textit{z}-direction and $\Gamma_l=2l\zeta/R_0$. Suppose the mode spirals back and forth along the \textit{z}-axis and is confined within two caustics located at $\pm z_c$, see Fig. 4, then $l\approx2\pi R(z_c)/\lambda$, where $\lambda$ is the wavelength of interest. $E(r,\varphi, z)$ can be separated as $E(r,\varphi,z)=\Psi_{lp}(r,z)Z_{lp}(z)\exp{(il\varphi)}$, where $\Psi_{lp}(r,z)$ is\cite{Sumetsky2004, Louyer2005}:
\begin{equation}
\Psi_{lp}(r,z)=\left\{ \qquad
\begin{aligned}A_lJ_l(k_\varphi r),& r<R(z)\\ H_l^{(2)}(\frac{k_\varphi r}{n})+B_lH_l^{(1)}(\frac{k_\varphi r}{n}), & r>R(z)\end{aligned}\right.
\label{eq2}
\end{equation}
while $Z_{lp}(z)$ is:
\begin{equation}
Z_{lp}(z)=C_{lp}H_p(\sqrt{\frac{\Gamma_l}{2}}z)exp(-\frac{\Gamma_l}{4}z^2).
\label{eq3}
\end{equation}
In Eq.~\ref{eq2}, $n$ is the refractive index of the BLMR material, $k_\varphi$ is defined as the azimuthal component of the propagation constant $k$, such that $k_\varphi=kR(z_c)/R(z)$. For a specific mode, $k$ is represented by $k_{lp}$. $J_l$ and $H_l^{(1,2)}$ are Bessel and Hankel functions, respectively. $A_l$ and $B_l$ are coefficients which can be derived by considering the boundary condition at $r=R(z)$. In Eq.~\ref{eq3}, $H_p$ is a Hermite function and $C_{lp}=[\Gamma_l/(\pi 2^{2p+1}(p!)^2)]^{\frac{1}{4}}$ is a coefficient.
\begin{figure}
\begin{center}
\includegraphics[width=75mm]{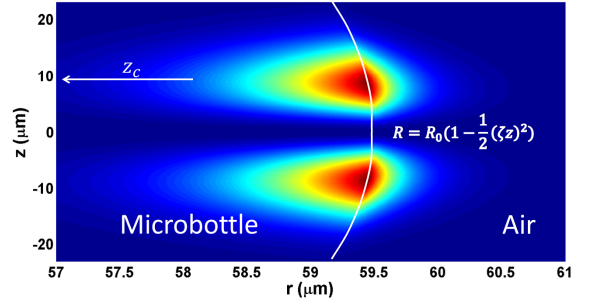}
\end{center}
\caption{Electromagnetic field distribution of the $(l,p)=(160,1)$ BLMR mode in cylindrical coordinates. $z_c$ is the caustic point.}
\label{sim}
\end{figure}

The mode volume of the fundamental BLMR mode, $p=1$, is the smallest. In our case, $R_0=59.5~\mathrm{\mu m}$, $\zeta\approx 0.0052 ~\mathrm{\mu m^{-1}}$ and we choose $(l,p)=(160,1)$ and $\pm z_c=10~\mathrm{\mu m}$. By substituting the above values into Eq.~\ref{eq1}-\ref{eq3}, the mode volume, $V_{160,1}$, is calculated to be about $15,000~\mathrm{\mu m^3}$ from Eq.\ref{eq4}. The mode distribution is plotted in Fig.~\ref{sim}. For comparison, the volume of the fundamental mode of a silica microsphere can be deduced in a similar way\cite{Braginsky1989}. For a microsphere with a radius of $60~\mathrm{\mu m}$, the mode volume is about $3500~\mathrm{\mu m^3}$. The larger mode volume in the BLMR corresponds to a Raman threshold which is more than double that in a microsphere as obtained in earlier works\cite{Agha2009, Spillane2002}.

\begin{figure}
	\begin{center}
	\includegraphics[width = 60mm]{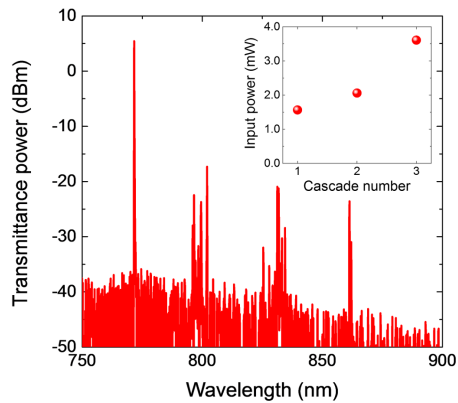}
	\end{center}
	\caption{Spectrum of cascaded Raman scattering. Pump light at $771.6~\mathrm{nm}$ generates  first-order Raman scattering around $802.0~\mathrm{nm}$, this generates the second--order around $831.2~\mathrm{nm}$ and this generates the third-order around $861.3~\mathrm{nm}$.}
	\label{f3}
\end{figure}

Finally, we demonstrate cascaded Raman lasing using the same hollow microresonator. The pump light source was changed to a wavelength of 771.6~nm (Newfocus Velocity 6712). A BLMR mode was found with a Q of more than $2\times10^7$. At a pump power greater than 1.5~mW, the first order Raman lasing peak at 802.0~nm appears. By increasing the pump power to more than 2~mW, the first-order Raman laser peak becomes strong enough to excite the second-order Raman peak at 831.2~nm. By such a cascaded process, a third-order Raman laser at 861.3~nm was excited when the pump laser power exceeded 3.8 mW. A spectrum containing three orders of cascaded Raman lasing is given in Fig.~\ref{f3}. In principle, such a process can continue with further increases to the input power. However, as shown by Min \textit{et al.}\cite{Min2003}, the cascaded Raman threshold increases at a cubic rate that is proportional to the Raman order number.  Hence, in our case, the fourth order threshold is beyond the maximum output of the laser source.

In conclusion, we have realized stimulated Raman scattering in BLMRs. Even though the bottle-like WGR has a larger mode volume compared with other WGRs, a Raman threshold of 1.6~mW is still achievable. Cascaded Raman lasing can also be obtained.
In the future, Raman lasing in the hollow BLMR could be utilized for high sensitivity sensing applications, such as aerostatic pressure sensing. Aerostatic pressure tuning of WGRs was previously demonstrated in solid microspheres\cite{Martin2013}, but the sensitivity was low. In hollow structures, such as microbubbles, high sensitivity and resolution can be set during fabrication, since the sensitivity improves with reduced wall thickness\cite{Henze2011, Yang2015}. However, very thin walls cannot withstand high pressure; therefore, there is a trade off between sensitivity and dynamic range. The Raman laser linewidth is much narrower than the natural linewidth of the passive cavity, so that, even with a thick wall, it is still possible to resolve the very small changes at high pressure. The BLMR in this work has an estimated\cite{Henze2011} sensitivity of $1.14~\mathrm{kHz/Pa}$ while the Raman laser linewidth is in KHz range, so such a device is feasible to implement an aerostatic pressure sensor with a good resolution working in high pressure range. Aside from the potential application to high pressure sensing, laser sensing in WGRs\cite{Li2014,Ozdemir2014} may also be realized in this structure.

This work was supported by the Okinawa Institute of Science and Technology Graduate University.

\end{document}